\documentclass[a4paper,12pt]{article}
\usepackage{amsmath}
\usepackage{amssymb}
\usepackage{latexsym}
\usepackage{graphicx}
\newcommand{\be}{\begin{eqnarray}}
\newcommand{\ee}{\end{eqnarray}}
\begin{document}
\begin{titlepage}

\begin{centering}
\vspace{.3in}
{\Large{\bf Energy distribution of the Einstein-Klein-Gordon system for a static spherically
symmetric spacetime in $(2+1)$-dimensions}}
\\

\vspace{.5in} {\bf  I. Radinschi$^*$\footnote{radinschi@yahoo.com},
Th. Grammenos$^\dag$}\footnote{thgramme@uth.gr}\\

\vspace{0.3in}
{\it $^*$ Physics Department, ``Gh. Asachi'' Technical University,\\
6600 Iasi, Romania\\
$^\dag$  Department of Mechanical \& Industrial Engineering, University of Thessaly,\\
383 34 Volos, Greece\\
}
\end{centering}

\vspace{0.7in}
\begin{abstract}
\par\noindent
We use M\o ller's energy-momentum complex in order to explicitly compute the energy and momentum density distributions
for an exact solution of Einstein's field equations with a negative cosmological constant minimally coupled to a static massless scalar field in a
static, spherically symmetric background in $(2+1)$-dimensions.
\end{abstract}

\end{titlepage}
\newpage
\numberwithin{equation}{section}
\section{Introduction}

One of the most interesting issues in General Relativity is the subject of\break
energy-momentum localization. This problem of defining in an acceptable
manner the energy-momentum density hasn't got a generally accepted answer yet. The
localization of energy implies the use of various energy-momentum complexes,
including the prescriptions of Einstein \cite{1}, Landau and Lifshitz \cite{2},
Papapetrou \cite{3}, Bergmann \cite{4} and Weinberg \cite{5}. All of these prescriptions have a
drawback: the calculations are restricted to quasi-Cartesian
coordinates. On the other hand, M\o ller \cite{6} proposed an energy-momentum
complex which enables one to calculate the energy distribution in any
coordinate system and not only in quasi-Cartesian coordinates. Hence, an
interesting question is whether any of the aforementioned prescriptions provides the best option for
energy-momentum localization and, of course, under what conditions.

The problem of energy-momentum localization by using the energy-momentum
complexes was revived by Virbhadra and his collaborators. He established an important
result showing that different
energy-complexes (ELLPW) yield the same result for a general non-static
spherically symmetric metric of the Kerr-Schild class \cite{7}. Furthermore, these
definitions (ELLPW) comply with the quasi-local mass definition of Penrose
for a general non-static spherically symmetric metric of the Kerr-Schild
class. However, these prescriptions disagree in the case of the most general
non-static spherically symmetric metric. According to Virbhadra,  although the
energy-momentum complexes behave under general coordinate
transformations like non-tensorial objects, the local conservation laws obtained by them hold in all
coordinate systems. Also, many interesting results recently obtained \cite{8}-\cite{9}
point out that the energy-momentum complexes are powerful tools for
obtaining the energy distribution in a given spacetime. Interesting works
were done with the energy-momentum complexes in $2$- and $3$-dimensional
spacetimes \cite{10}. In this context, of great importance is the Cooperstock hypothesis \cite{11} which
states that energy and momentum are confined to the regions of
non-vanishing energy-momentum tensor for the matter and all non-gravitational
fields.

Chang, Nester and Chen \cite{12} showed that the energy-momentum
complexes are actually quasilocal and give a legitimate expression for the
energy-momentum. Further, they concluded that there exists a direct relationship
between energy-momentum complexes and quasilocal expressions, since every
energy-momentum complex is associated with a legitimate Hamiltonian
boundary term.

Regarding M\o ller's energy-momentum complex, there is a plethora of
results \cite{13}-\cite{14} that recommend it as an efficient tool for the localization of
energy. According to Lessner \cite{15}, M\o ller's energy-momentum
complex is significant for discribing the concepts of energy and momentum in General Relativity.
In fact, he stated that ``The energy-momentum
four-vector can transform according to special relativity only if it is
transformed to a reference system with an everywhere constant velocity. This
cannot be achieved by a global Lorentz transformation''.

For Bondi \cite{16} ``a nonlocalizable form of
energy is not admissible in General Relativity, because any form of energy
contributes to gravitation and so its location can in principle be
found''.

Based on the above works and conclusions we have decided to use M\o ller's energy-momentum
complex for evaluating the energy distribution of an exact solution for
a static massless scalar field minimally coupled to gravity in $(2+1)$ dimensions.

Consequently, in this work we have implemented M\o ller's formalism to calculate the energy
distribution for a geometry given recently by Daghan and Bilge \cite{17}. They used a series of transformations to reduce
to second order the system of differential equations for the Einstein's field equations with a negative
cosmological constant, minimally coupled to a massless scalar field in a static,
spherically symmetric background in $(2+1)$-dimensions. In this way, the authors obtained an
exact solution in a form that generalizes the AdS metric as given by Matschull \cite{18}.
Throughout we use geometrized units, i.e. $c = G = 1$,
while $g_{\mu\nu}$ has the signature $(-, +, +)$, Greek indices run from $0$ to $2$ and Latin indices
run over the spatial coordinate values.
The rest of the paper is organized as follows. In Sec.2 we give a short review of the concept of energy-momentum
complexes and present M\o ller's prescription. In Sec.3 we briefly present the $(2+1)$-dimensional, spherically symmetric
background used, while in Sec.4 we explicitly compute the energy and momentum density distributions of the Einstein-Klein-Gordon system using
M\o ller's energy-momentum complex. Finally, in Sec.5 a discussion of the obtained results and some concluding remarks are given.

\section{Energy-Momentum Complexes and\\ M\o ller's Prescription}
The conservation laws of energy and momentum  for an isolated,
i.e., no external force is acting on the system, physical system in
the Theory of Special Relativity are expressed by a set of
differential equations. Defining $T^{\mu}_{\nu}$ as the symmetric
energy-momentum tensor of matter and of all non-gravitational
fields, the conservation laws are given by
\begin{equation}\label{2.1}
T^{\mu}_{\nu,\, \mu} \equiv \frac{\partial T^{\mu}_{\nu}}{\partial
x^{\mu}}=0,
\end{equation}
where
\begin{equation}\label{2.2}
\rho=T^{t}_{t}\hspace{1cm}j^{i}=T^{i}_{t}\hspace{1cm}p_{i}=-T^{t}_{i}
\end{equation}
are the energy density, the energy current density, and the
momentum density, respectively.
Making the transition from Special to General Theory of
Relativity, one adopts a simplicity principle which is called
principle of minimal gravitational coupling. As a result,
the conservation equation is now written as
\begin{equation}\label{2.3}
T^{\mu}_{\nu;\, \mu} \equiv
\frac{1}{\sqrt{-g}}\frac{\partial}{\partial
x^{\mu}}\left(\sqrt{-g}\,T^{\mu}_{\nu}\right)-\Gamma^{\kappa}_{\nu\lambda}T^{\lambda}_{\kappa}=0,
\end{equation}
where $g$ is the determinant of the metric tensor
$g_{\mu\nu}(x)$. The conservation equation may also be written as
\begin{equation}\label{2.4}
\frac{\partial}{\partial
x^{\mu}}\left(\sqrt{-g}\,T^{\mu}_{\nu}\right)=\xi_{\nu},
\end{equation}
where
\begin{equation}\label{2.5}
\xi_{\nu}=\sqrt{-g}\Gamma^{\kappa}_{\nu\lambda}T^{\lambda}_{\kappa}
\end{equation}
is a non-tensorial object. For $\nu=t$ this means that the
matter energy is not a conserved quantity for the physical
system\footnote{It is possible to restore the conservation law by
introducing a local inertial system for which at a specific
spacetime point $\xi_{\nu}=0$ but this equality by no means holds
in general.}. From a physical point of view, this lack of energy
conservation can be understood as the possibility of transforming
matter energy into gravitational energy and vice versa. However,
this remains an open problem and it is widely believed that in
order to solve it one has to take into account the gravitational
energy.

By a well-known procedure, the non-tensorial object $\xi_{\nu}$
can be written  as
\begin{equation}\label{2.6}
\xi_{\nu}=-\frac{\partial}{\partial
x^{\mu}}\left(\sqrt{-g}\,\vartheta^{\mu}_{\nu}\right),
\end{equation}
where
$\vartheta^{\mu}_{\nu}$ are certain functions of the metric tensor
and its first order derivatives. Therefore, the energy-momentum
tensor of matter $T^{\mu}_{\nu}$ is replaced by the expression
\begin{equation}\label{2.7}
\theta^{\mu}_{\nu}=\sqrt{-g}\left(T^{\mu}_{\nu}+\vartheta^{\mu}_{\nu}\right),
\end{equation}
which is called energy-momentum complex since it is a
combination of the tensor $T^{\mu}_{\nu}$ and a pseudotensor
$\vartheta^{\mu}_{\nu}$ describing the energy and  momentum of the
gravitational field. The energy-momentum complex satisfies a
conservation law in the ordinary sense, i.e.,
\begin{equation}\label{2.8}
\theta^{\mu}_{\nu,\, \mu}=0
\end{equation}
and it can be written as
\begin{equation}\label{2.9}
\theta^{\mu}_{\nu}=\chi^{\mu\lambda}_{\nu\,\,\,,\lambda},
\end{equation}
where
$\chi^{\mu\lambda}_{\nu}$ are called superpotentials and are
functions of the metric tensor and its first order derivatives.

It is evident that the energy-momentum complex is not uniquely
determined by the condition that its usual divergence is zero
since a quantity with an identically vanishing divergence can
always be added to the energy-momentum complex.

The energy-momentum complex of M{\o}ller in a  four-dimensional
background is given as \cite{6}
\begin{equation}\label{2.10}
\mathcal{J}^{\mu}_{\nu}=\frac{1}{8\pi}\xi^{\mu\lambda}_{\nu\,\,
,\, \lambda},
\end{equation}
where M{\o}ller's superpotential $\xi^{\mu\lambda}_{\nu}$ is of the form
\begin{equation}\label{2.11}
\xi^{\mu\lambda}_{\nu}=\sqrt{-g} \left(\frac{\partial
g_{\nu\sigma} }{\partial x^{\kappa} }- \frac{\partial
g_{\nu\kappa}}{\partial x^{\sigma}
}\right)g^{\mu\kappa}g^{\lambda\sigma},
\end{equation}
with the antisymmetric
property
\begin{equation}\label{2.12}
\xi^{\mu\lambda}_{\nu}=-\xi^{\lambda\mu}_{\nu}.
\end{equation}

It is easily seen that  M{\o}ller's energy-momentum complex
satisfies the local conservation equation
\begin{equation}\label{2.13}
\frac{\partial \mathcal{J}^{\mu}_{\nu}}{\partial x^{\mu}}=0,
\end{equation}
where $\mathcal{J}^{0}_{0}$ is the energy density and
$\mathcal{J}^{0}_{i}$ are the momentum density components.

Thus, in M{\o}ller's prescription the energy and momentum for a
four-dimensional background are given by
\begin{equation}\label{2.14}
P_{\mu}=\int\int\int \mathcal{J}^{0}_{\mu}dx^{1}dx^{2}dx^{3}.
\end{equation}
Specifically the energy of the physical system in a
four-dimensional background is
\begin{equation}\label{2.15}
E=\int\int\int
\mathcal{J}^{0}_{0}dx^{1}dx^{2}dx^{3}.
\end{equation}

It should be noted that the calculations are not anymore
restricted to quasi-Cartesian coordinates but can be utilized in
any coordinate system.

\section{The $(2+1)$-dimensional Static Spherically\\ Symmetric Background}
The line element of a three-dimensional static, spherically symmetric, Lorentzian spacetime is given as
\begin{equation}\label{3.1}
ds^2=-e^{2f(r)}dt^2+e^{2f(r)}dr^2+e^{-2h(r)}d\theta^2,
\end{equation}
in $(t,r,\theta)$ coordinates. Einstein's field equations with a cosmological constant $\Lambda$ are
\begin{equation}\label{3.2}
R_{\mu\nu}-\frac{1}{2}g_{\mu\nu}R+\Lambda g_{\mu\nu}=\kappa T_{\mu\nu},
\end{equation}
where
\begin{equation}\label{3.3}
T_{\mu\nu}=\partial_{\mu}\phi\partial_{\nu}\phi-\frac{1}{2}g_{\mu\nu}(\partial_{\lambda}\phi\partial^{\lambda}\phi)
\end{equation}
is the energy-momentum tensor of a massless scalar field $\Phi$ \cite{19}
satisfying the wave equation:
\begin{equation}\label{3.4}
g^{\mu\nu}\nabla_{\mu}\nabla_{\nu}\phi=(-g)^{-\frac{1}{2}}\partial_{\mu}[(-g)^{-\frac{1}{2}}g^{\mu\nu}\partial_{\nu}\phi].
\end{equation}

The above wave equation of the Klein-Gordon type can be integrated giving for the first derivative of the static scalar field $\phi$:
\begin{equation}\label{3.5}
\phi^\prime=\frac{\lambda}{\sqrt{2\pi}}e^{h(r)},
\end{equation}
where $\lambda$ is an integration constant.
By using a series of transformations and taking $\Lambda = -1$, Daghan and Bilge \cite{17} managed to solve analytically
the above Einstein-Klein-Gordon system of
differenial equations for $f(r)$, $h(r)$ and $\phi$ and obtained the following form for the line element:
\begin{equation}\label{3.6}
ds^2=-\left[\frac{\cosh \chi}{\sinh \chi}\right]^{\mu}\cosh \chi\sinh \chi \,d\tilde{t}^2+
d\chi^2+(c^2+4\lambda^2)\left[\frac{\cosh\chi}{\sinh\chi}\right]^{-\mu}\cosh\chi\sinh\chi\, d\tilde{\theta}^2,
\end{equation}
where $\chi$, $\tilde{t}$, $\tilde{\theta}$ are defined through the following transformations:
\begin{equation}\label{3.7}
\chi=-\frac{1}{2}\ln (\tan\frac{\varphi}{2}),
\end{equation}
\begin{equation}\label{3.8}
\tilde{t}=\frac{\rho_0}{2}t,
\end{equation}
\begin{equation}\label{3.9}
\tilde{\theta}=\frac{2}{\rho_0}\theta.
\end{equation}
$c\neq 0$ being an integration constant and $\mu = \displaystyle{\frac{c}{\sqrt{c^2 +4\lambda^2}}}$.\\
$\varphi\in (0, \pi /2)$ and it can be seen that for $\varphi =\pi /2$ a curvature singularity arises.
Furthermore, $\varphi$ is related to $f(r)$ and $h(r)$ through
\begin{equation}\label{3.10}
e^{f(r)}=\frac{1}{2}\rho\cos\varphi,
\end{equation}
\begin{equation}\label{3.11}
e^{h(r)}=\frac{1}{\sqrt{c^2+2\lambda^2}}\rho\sin\varphi,
\end{equation}
while
\begin{equation}\label{3.12}
\rho(\varphi)=\frac{\rho_0}{\sqrt{2}}(1+\sin\varphi)^{\frac{\mu}{2}}(\sin\varphi)^{-\frac{1}{2}}(\cos\varphi)^{\frac{-1-\mu}{2}}
\end{equation}
and $\rho_0$ is an integration constant too.

It should be noted, that for $\lambda$ = 0 and c = 1 one gets the three-dimensional Anti-deSitter solution given by Matschull in \cite{18}.

\section{Energy and Momentum Density Distributions}
We first have to evaluate the superpotentials in the context of M\o ller's prescription for the
spacetime described by the line element (\ref{3.6}).
There are four non-zero independent superpotentials:
\begin{equation}\label{4.1}
\begin{split}
\xi_0^{01} &=(-\mu+\cosh(2\chi))\sqrt{c^2+4\lambda^2},\\
\\
\xi_0^{10} &=-\xi_{0}^{01}=(\mu-\cosh(2\chi))\sqrt{c^2+4\lambda^2},\\
\\
\xi_2^{21} &=\frac{(9+4\lambda^2)(\mu+\cosh(2\chi))}{\sqrt{c^2+4\lambda^2}},\\
\\
\xi_2^{12} &=-\xi_{2}^{21}=-\frac{(9+4\lambda^2)(\mu+\cosh(2\chi))}{\sqrt{c^2+4\lambda^2}}.
\end{split}
\end{equation}
Substitution of the above superpotentials into eq.(\ref{2.10}) yields the energy density distribution
\begin{equation}\label{4.2}
\mathcal{J}_0^0=\frac{4}{K}\sqrt{c^2+4\lambda^2}\cosh \chi\sinh \chi,
\end{equation}
where $K$ is an integration constant, while the momentum density distributions obtained are
\begin{equation}\label{4.3}
\mathcal{J}_1^0=0,
\end{equation}
\begin{equation}\label{4.4}
\mathcal{J}_2^0=0.
\end{equation}
Replacing equations (\ref{4.3}-\ref{4.4}) into eq. (\ref{2.14}) yields a zero momentum, while by substituting eq.(\ref{4.2}) into eq.(\ref{2.15}),
we get the energy of a massless scalar field contained in
a ``sphere'' of radius $\mathcal{X}$ for the three-dimensional spacetime described by the line element(\ref{3.6})
\begin{equation}\label{4.5}
E(\mathcal{X})=\frac{4\pi}{K}\sqrt{c^2+4\lambda^2}(\cosh(2\mathcal{X})-1).
\end{equation}
This is also the energy of the gravitational field that the scalar field experiences at a finite distance $\mathcal{X}$. Thus, the energy
given by equation (\ref{4.5}) is also called effective gravitational mass ($E = M_{\text{eff}}$) of the spacetime under study.
The energy content of an infinitely large ``sphere'' is infinite, as it is also evident from Figure 1.
\vspace{1cm}
\begin{figure}[h!]
\includegraphics[scale=1]{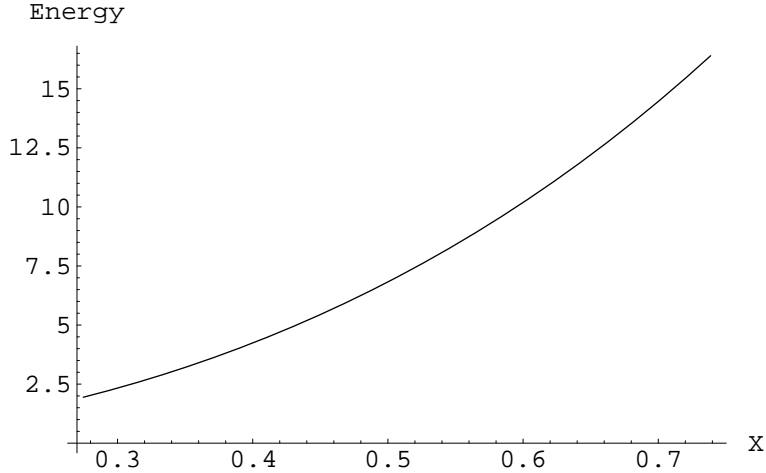}
\caption{Energy versus distance}
\end{figure}
\newpage

\section{Discussion - Conclusions}
As we pointed out previously, the subject of the localization of energy continues to be open,
since Einstein, in his theory of special relativity, declared that mass is equivalent to energy.
Misner {\it et al.} \cite{20} concluded that the energy is localizable only for spherical systems.
On the other hand, Cooperstock and Sarracino \cite{21} demonstrated that if the energy is localizable in spherical systems
then it is also localizable in any spacetime.

Now, concerning the use of several energy-momentum complexes for the localization of energy and momentum, we remark
that although it has many adepts, there has also been a great deal of critisism related to the use of such
prescriptions. The main disadvantage of the energy-momentum complexes lies in that most of them restrict
one to make calculations in quasi-Cartesian coordinates. Nevertheless, M\o ller's energy-momentum complex
is the only one allowing calculations in any coordinate system.

In this work, we have explicitly calculated the energy and momentum density distributions associated with an exact solution of
Einstein's field equations with a negative cosmological constant and minimally coupled to a static massless scalar field in a
static, spherically symmetric, three-dimensional spacetime. The energy and momentum distributions are computed
by use of M\o ller's energy-momentum complex.

A physically acceptable expression for the energy density distribution is obtained. The momentum density distributions
for the used $(2+1)$-dimensional, static spherically symmetric background vanish. Additionally, the effective gravitational mass,
i.e. the total energy,
is explicitly evaluated for the specific gravitational background. We have shown that the energy distribution
can be interpreted as the effective gravitational mass corresponding to the $(2+1)$-dimensional background.
The parameter $\mu \in (0,1)$ appearing in (\ref{4.5}) expresses the strength
of the coupled scalar field. One can see that for the special case $\mu = 1$, the three-dimensional Anti-deSitter spacetime is obtained.

Furthermore, M\o ller's energy-momentum complex applied to this $(2+1)$-dimensional solution satisfies
the local conservation equation (\ref{2.13}).

We point out that M\o ller's formalism provides physically satisfactory results for the aforementioned background. Finally,
these results support Lessner's viewpoint \cite{15} about the significance of M\o ller's energy-momentum complex as
a powerful tool for the evaluation of the energy and momentum density distributions in a given geometry.

\section*{Acknowledgements}
The authors are indebted to Dr. E.C. Vagenas for useful suggestions and comments.


\end{document}